\documentclass[10pt,twoside,letter]{article}
\usepackage{amsmath,amssymb,multicol,mathrsfs,graphicx}
\AtBeginDvi{}
\usepackage[%
linkcolor=blue,%
citecolor=red,%
bookmarks=true,%
bookmarksnumbered=true,%
colorlinks=true,%
dvipdfm%
]{hyperref}

\title{Fourier Analysis of the Parametric Resonance of the Neutrino Oscillation
  in the Presence of Inhomogeneous Matter}

\author{ %
  Joe Sato$^{\textrm{a}}$
  \thanks{email: joe@phy.saitama-u.ac.jp},
  Masafumi Koike$^{\textrm{a}}$
  \thanks{email: koike@krishna.th.phy.saitama-u.ac.jp},
  Toshihiko Ota$^{\textrm{b}}$
  \thanks{email: Toshihiko.Ota@physik.uni-wuerzburg.de},
  and
  Masako Saito$^{\textrm{a}}$
  \thanks{email: msaito@krishna.th.phy.saitama-u.ac.jp}
  \\
  \textit{\small{
  $^{\textrm{a}}$
  Physics Department, Saitama University,
  255 Shimo-Okubo, 
  Sakura-ku, Saitama 338-8570, Japan
  }
  }
  \\
  \textit{\small{
  $^{\textrm{b}}$
  Instit\"{u}t f\"{u}r Theoretische Physik und Astrophysik,
  Universit\"{a}t W\"{u}rzburg,
  97074 W\"{u}rzburg, Germany
  }
  }
}

\date{}

\begin{document}

\maketitle

\begin{abstract}
  We study the parametric resonance of the neutrino oscillation through the
  matter whose density varies spatially.
  The Fourier analysis of the matter effect enables us to clarify the parametric
  resonance condition, which is summarized in a frequency matching between the
  neutrino oscillation and the spatial variation of the matter density.
  As a result, the $n$-th Fourier mode of a matter density profile modifies the
  energy spectrum of the $\nu_{\mu} \to \nu_{\textrm{e}}$ appearance
  probability at around the $n$-th dip.
\end{abstract}

\vspace{2.0\baselineskip}

The oscillation probability of neutrinos passing through the interior of the
Earth is strongly affected by the matter on the baseline.
In such cases, the spatial variation of the matter density can make the
probability quite different from that under the constant matter
density~\cite{ParametricResonance,Petcov,Ota}.
The possibility of the parametric resonance of the neutrino oscillation with the
matter density has been discussed~\cite{ParametricResonance}.
We exploit the Fourier analysis of the matter density profile to study the
effect of parametric resonance~\cite{Ota,Koike:1998hy}.
Generally, the parametric resonance of an oscillation emerges when an
oscillation parameter changes with a resonant frequency, which is typically
twice the natural frequency of the system.
Our formulation based on the Fourier analysis gives a clear view of 
this frequency matching, which is essential to the parametric resonance.

\begin{figure}
  \hfill
    \includegraphics[height=38mm]{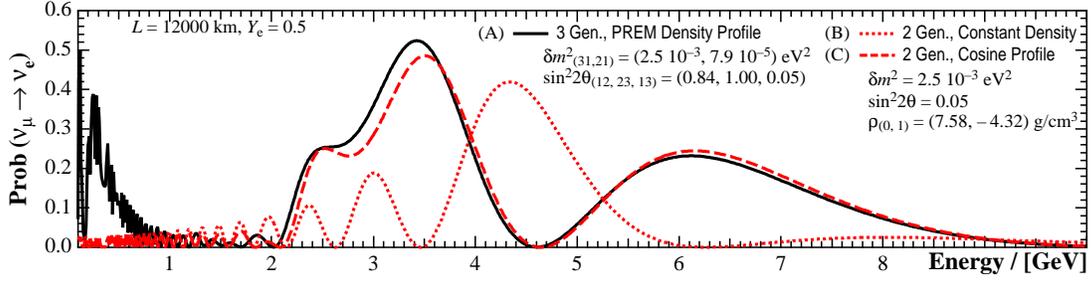}
  \hfill{}
  \caption{ %
    Oscillation probability from $\nu_{\mu}$ to $\nu_{\textrm{e}}$
    for the baseline length of $12000 \, \mathrm{km}$.
    The solid curve (A) is evaluated by three-flavor analysis with
    the matter density profile taken from the PREM.
    The dotted (B) and broken (C) curves are results of the two
    flavor calculation with (B) constant density with $\rho_{0} = 7.58\,
    \mathrm{g/cm^{3}}$, and (C) cosine profile $\rho(x) = \rho_{0} + \rho_{1}
    \cos 2\pi x/L$ with $\rho_{1} = - 4.32\, \mathrm{g/cm^{3}}$.
    Values of other parameters taken are described in the figure.
  }
  \label{fig:3Gen-vs-2Gen}
\end{figure}
We carry out our analysis in a two-flavor framework.
This simplification does not spoil our point as can be seen in
Fig.~\ref{fig:3Gen-vs-2Gen}, in which we compare the $\nu_{\mu} \to
\nu_{\textrm{e}}$ appearance probability in the three- and
two-flavor calculation, taking the baseline length $L = 12000 \, \mathrm{km}$.
 The three-flavor result assumes the density profile of 
 the Preliminary Reference Earth Model (solid curve), 
 while two-flavor result uses the constant density case (dotted curve) 
 and the cosine variation around the average density (broken curve). 
 Renormalized by the factor of $\sin^{2}2\theta_{23}$,
 the two-flavor result with the cosine profile shows the similar feature 
 as the three-flavor one.

 The evolution equation of two flavors of neutrinos $\nu_{\textrm{\scriptsize
     e}}$ and $\nu_{\mu}$ in the matter density $\rho(x)$ is 
\begin{equation}
  \mathrm{i} \frac{\mathrm{d}}{\mathrm{d} x}
  \begin{pmatrix} \nu_{\textrm{e}}(x) \\ \nu_{\mu}(x) \end{pmatrix}
  =
  \frac{1}{2E}
  \biggl[
    \frac{\delta m^{2}}{2}
    \begin{pmatrix}
      - \cos 2\theta & \sin 2\theta \\ \sin 2\theta & \cos 2\theta
    \end{pmatrix}
    +
    \begin{pmatrix} a(x) & 0 \\ 0 & 0 \end{pmatrix}
  \biggr]
  \begin{pmatrix} \nu_{\textrm{e}}(x) \\ \nu_{\mu}(x) \end{pmatrix}
  \, ,
  \label{eq:evolution_eq}
\end{equation}
where $\delta m^{2}, \theta$, and $E$ are the mass-square difference, the mixing
angle, and the energy of neutrinos, respectively.
The matter effect $a(x)$ is given by 
\begin{math}
  a(x) = 2\sqrt{2} E G_{\textrm{F}} N_{\textrm{A}}
       Y_{\textrm{e}} \rho(x),
\end{math}
where $G_{\textrm{F}}$ is the Fermi constant, 
$N_{\textrm{A}}$ is the Avogadro number, 
and $Y_{\textrm{e}}$ is the proton-to-nucleon ratio. 
Defining
$\Delta \equiv \delta m^{2} L/2E$,
$\Delta_{\textrm{m}}(x) \equiv a(x)L/2E$, and
\begin{math}
  z(x) =
  \nu_{\textrm{e}}(x)
  \exp [
    \int_{0}^{x} \mathrm{d}s \,
    \mathrm{i} \Delta_{\textrm{m}}(s)/2
  ],
\end{math}
and introducing an Fourier expansion
\begin{math}
  \Delta_{\textrm{m}}(\xi)
  = \sum_{n = 0}^{\infty}\Delta_{\textrm{m}n } \cos 2 n \pi \xi
\end{math}
with $\xi = x / L$, we obtain from Eq.~(\ref{eq:evolution_eq})
\begin{equation}
\begin{split}
  z''(\xi)
  &
  + \frac{1}{4} \Bigl[ 
    \bigl( \Delta_{\textrm{m} 0} - \Delta \cos 2\theta \bigr)^{2}
      + \Delta^{2} \sin^{2} 2\theta
      + 2 ( \Delta_{\textrm{m} 0} - \Delta \cos 2 \theta )
        \sum_{n = 1}^{\infty} \Delta_{\textrm{m} n} \cos 2n \pi \xi
    \\ &
    + \Bigl(
        \sum_{n = 1}^{\infty} \Delta_{\textrm{m} n} \cos 2n \pi \xi
      \Bigr)^{2}
    + 4n\pi \mathrm{i} 
      \sum_{n = 1}^{\infty} \Delta_{\textrm{m} n} \sin 2n \pi \xi
  \Bigr] z(\xi)
  = 0 \, .
\end{split}
\label{eq:diff-eq-z-2}
\end{equation}

\begin{figure}
  \hfill
    \includegraphics[height=40mm]{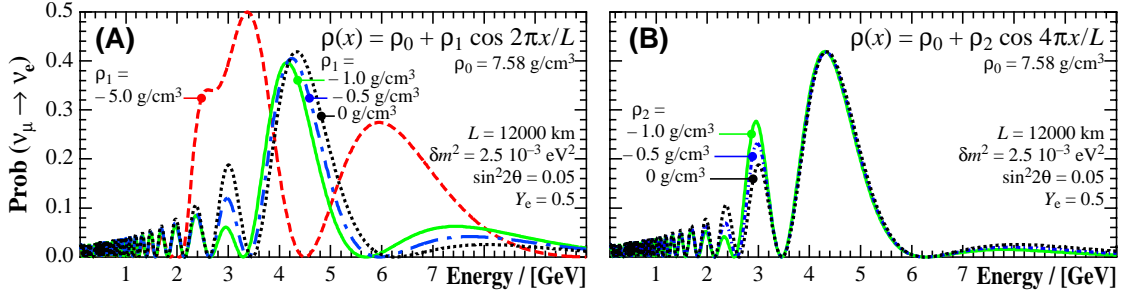}
  \hfill{}
  \caption{ %
    The $\nu_{\mu} \to \nu_{\textrm{e}}$ oscillation probabilities
    with $\rho_{n}= 0, -0.5,$ and $-1.0 \, \mathrm{g/cm^{2}}$, where
    (A) $n = 1$ and (B) $n = 2$.
    The former includes $\rho_{1} = -5.0 \, \mathrm{g/cm^{2}}$ case also.
    Values of other parameters are shown in the graphs.
  }
  \label{fig:rhon-effect}
\end{figure}

The parametric resonance can arise under the presence of the density profile.
Suppose that only the $n$-th Fourier mode of the density profile is present so
that
\begin{math}
  \Delta_{\textrm{m}}(\xi)
  =
  \Delta_{\textrm{m}0}
  + \Delta_{\textrm{m}n} \cos 2n\pi\xi.
\end{math}
The parametric resonance is expected when the frequency of the density profile
$2n \pi$ is twice 
the natural frequency of the system
\begin{math}
  \omega_{0}
  \equiv
  \sqrt{
    ( \Delta_{\textrm{m} 0} - \Delta \cos 2\theta )^{2}
    + \Delta^{2} \sin^{2} 2\theta
     -\frac{1}{2}\sum_{n=1}^{\infty}  \Delta_{\textrm{m} n}^{2}
  } / 2 \, ,
  \label{eq:freq-constant-matter}
\end{math}
\textit{i.e.} $\omega_{0} = 2n \pi / 2 = n \pi$.
This frequency matching condition amounts to
\begin{equation}
  E =
  E^{\textrm{(res)}}_{n} \equiv
  \frac{\delta m^{2} L}{2}
  \frac{1}{
    \Delta_{\textrm{m} 0} \cos 2\theta
    \pm \sqrt{
      4 n^{2} \pi^{2}
      - \Delta_{\textrm{m} 0}^{2} \sin^{2} 2\theta
      -\frac{1}{2}\sum_{n=1}^{\infty}  \Delta_{\textrm{m} n}^{2}
    }
  } \, .
\label{eq:para_reso_energy}  
\end{equation}
We show in Fig.~\ref{fig:rhon-effect} the $\nu_{\mu} \to
\nu_{\textrm{e}}$ oscillation probability for various values (A) of
$\rho_{1}$ and (B) of $\rho_{2}$.
Figure \ref{fig:rhon-effect} indicates that the $n$-th Fourier mode of the
matter density modifies the oscillation probability around the $n$-th dip:
$\omega_{0} = n\pi$, which corresponds to the parametric resonance energy
Eq. (\ref{eq:para_reso_energy}).
This modification is understood by the parametric resonance as shown
elsewhere~\cite{KOSS}.

In summary, we studied the parametric resonance of the neutrino oscillation
through the matter whose density varies spatially, analyzing the matter effect
by the Fourier expansion.
We have shown that the $n$-th Fourier mode of the density profile 
modifies the oscillation probability around the parametric resonance energy 
$E^{\textrm{(res)}}_{n}$ defined in
Eq.~(\ref{eq:para_reso_energy}).
This condition is understood in terms of the frequency matching between 
the natural frequency of the system and that of the density profile.

\end{document}